\documentclass[11pt,a4paper]{article}
\pdfoutput=1 % if your are submitting a pdflatex (i.e. if you have images in pdf, png or jpg format)
\usepackage{jheppub}	% jheppub includes hyperref,color, natbib, amsmath, amssymb, epsfig, graphicx
\usepackage{tikz}
\usetikzlibrary{positioning,arrows}
\usetikzlibrary{decorations.pathmorphing}
\usetikzlibrary{decorations.markings}
\usetikzlibrary{snakes}
\usetikzlibrary{matrix}

\usepackage{amsmath}
\usepackage{bm}
\usepackage[utf8]{inputenc}
\usepackage[english]{babel}
\usepackage{makeidx}
\usepackage{tikz}
\tikzset{every picture/.style={line width=0.75pt}} %set default line width to 0.75pt
\usepackage{amsfonts}
\usepackage{enumerate}
\usepackage{mathrsfs}
\usepackage{tensor}
\usepackage[autostyle]{csquotes}
\usepackage{subfig}

\def\be{\begin{equation}}
\def\ee{\end{equation}}
\def\bea{\begin{eqnarray}}
\def\eea{\end{eqnarray}}
\newcommand{\nn}{\nonumber}

\newcommand{\ft}[2]{{\textstyle\frac{#1}{#2}}}

\def\apjl{\ref@jnl{ApJ}}

\newcommand{\RNum}[1]{\uppercase\expandafter{\romannumeral #1\relax}}

\usepackage{amsmath,amssymb}
\usepackage{latexsym}
\usepackage{graphicx}
\usepackage{slashed}

 \usepackage[vcentermath,enableskew]{youngtab}
\usepackage{epsfig}

\def\be{\begin{equation}}
\def\ee{\end{equation}}
\def\bea{\begin{eqnarray}}
\def\eea{\end{eqnarray}}

\title{A note on rank 5/2 Liouville irregular block,\\ Painlev\'e \RNum{1} and
the ${\cal H}_0$ Argyres-Douglas theory}

\author[]{Hasmik Poghosyan and}
\author[]{Rubik Poghossian}
\affiliation[]{Yerevan Physics Institute \\
	Alikhanian Br. 2, 0036 Yerevan, Armenia}

\abstract{We study 4d type ${\cal H}_0$ Argyres-Douglas theory in $\Omega$-background 
by constructing Liouville irregular state of rank 5/2. The results are compared with 
generalized Holomorphic anomaly approach, which provides order by order expansion in 
$\Omega$-background parameters $\epsilon_{1,2}$. Another crucial test of our results 
provides comparison with respect to Painlev\'{e} \RNum{1} $\tau$-function, which was expected to 
hold in self-dual case $\epsilon_1=-\epsilon_2$. We also discuss Nekrasov-Shatashvili 
limit $\epsilon_1=0$, accessible either by means of deformed Seiberg-Witten curve, or 
WKB methods.
}
 \clearpage
\begin{document}
\tikzset{
line/.style={thick, decorate, draw=black,}
 }

\maketitle

%\flushbottom

\section{Introduction}

In famous paper \cite{Argyres:1995jj}, based on analysis of 
Seiberg-Witten curve  \cite{Seiberg:1994rs,Seiberg:1994aj}, the authors 
have shown, that there are specific points in moduli space of certain ${\cal N}=2$
supersymmetric gauge theories where electrically and magnetically charged 
particles simultaneously become massless. Since then such theories, commonly 
referred as Argyres-Douglas theories, attract non diminishing attention due to their 
rich physical content. Involvement of localization technique for investigation 
of  ${\cal N}=2$ SYM was instrumental for recent fascinating developments in this area 
\cite{Nekrasov:2002qd,Flume:2002az,Bruzzo:2002xf,Nekrasov:2003rj,Flume:2004rp}. 
The main idea was introduction of a specific gravitational background (so called 
$\Omega$-background) such that QFT path integrals for supersymmetry protected 
quantities get localized around discrete set of configurations and become 
explicitly calculable. One recovers initial SW theory in flat space time 
simply sending $\Omega$-background parameters (traditionally denoted by $\epsilon_1$ and 
$\epsilon_2$) to zero. Later developments have shown however, that keeping
background parameters finite is of great interest too.       
In particular a remarkable relationship between 2d CFT correlation functions and
partition function of gauge theory was uncovered 
\cite{Gaiotto:2009we,Alday:2009aq,Poghossian:2009mk}. 
In \cite{Gaiotto:2009ma,Gaiotto:2012sf} a systematic method is developed 
for constructing a new class of states (called rank $r$ irregular states) in 2d Liouville CFT 
by colliding insertion points of $r+1$ primary fields. Then it was shown that the corresponding 
gauge counterparts in case of $r=2$ and $r=3$ are just the AD theories denoted by ${\cal H}_1$ and ${\cal H}_2$ 
respectively. Such relationships were subject of further detailed investigation in 
\cite{Nishinaka:2012kn,Kanno:2013vi,Nishinaka:2019nuy,Kimura:2020krd,
Bonelli:2021uvf,Bonelli:2022ten,Kimura:2022yua,Consoli:2022eey,Fucito:2023plp}.
In a parallel development it was shown in \cite{Bonelli:2016qwg} that ${\cal H}_1$ and ${\cal H}_2$ partition  
functions are closely related to third and forth  Painlev\'{e} $\tau$-functions provided 
$\Omega$-background parameters are subject to condition $\epsilon_1=-\epsilon_2$.  
Unfortunately the simplest AD theory ${\cal H}_0$, which was expected to be related 
to Painlev\'{e} \RNum{1},  failed to be included coherently into 
above scheme until now, since a 2d CFT description was missing. The main purpose of 
current paper is to fulfill this gap. Namely, we have defined and carefully 
investigated irregular states of half-integer rank $r=5/2$ in Liouville theory (in fact our method can be 
easily generalized to other half-integer ranks as well). We have shown that this state, after 
separating a non-trivial factor, can be represented as a power series with terms
that are certain generalized descendants of rank $2$ irregular 
state. The higher level terms are recursively recovered, starting from the 
leading term, which simply coincides with rank $2$ state. Our proposal is that the overlap 
of rank $5/2$ irregular state with vacuum is the 2d CFT counterpart of partition function
for a slightly generalized version of ${\cal H}_0$ AD theory in general $\Omega$-background.
Given SW curve description of a theory, there is a powerful method to compute corrections 
in  $\Omega$-background parameters $\epsilon_{1,2}$ based on holomorphic anomaly recursion
\cite{Bershadsky:1993ta, Huang:2006si,Huang:2009md,Huang:2011qx,Huang:2013eja,Krefl:2010fm}
(see also \cite{Fucito:2023plp,Fucito:2023txg} for applications in various AD theories). 
This approach in some sense is complementary to irregular state method, since it provides  
exact expressions in coupling constant, but $\epsilon_{1,2}$ corrections are computed order 
by order.  The irregular state method does exactly the opposite. 

Using holomorphic anomaly recursion we have found exact formulae for the prepotential 
of our generalized ${\cal H}_0$ theory up to order $8$ in $\epsilon_{1,2}$. 
The results perfectly match with those obtained through irregular state computation. 
We also provide an additional consistency check by considering Nekrasov-Shatashvili limit 
$\epsilon_1=0$ and applying WKB analysis. A detailed investigation of the NS limit can 
be found in \cite{Ito:2018hwp}. 

In \cite{Gamayun:2012ma,Gamayun:2013auu}
the authors have found remarkable relations between  Painlev\'{e} 
$\RNum{6}$, $\RNum{5}$, $\RNum{3}_1$, $\RNum{3}_2$, $\RNum{3}_3$ $\tau$-functions
and $SU(2)$ gauge theory partition functions with ${\cal N}_f=4,3,2,1,0$  
hypermultiplets respectively, provided $\Omega$-background is restricted to 
$\epsilon_1=-\epsilon_2$. As already mentioned, in a similar 
manner the AD theories ${\cal H}_1$ and ${\cal H}_2$ are related to Painlev\'{e}  $\RNum{2}$ and $\RNum{4}$ 
\cite{Bonelli:2016qwg,Nishinaka:2019nuy}.   
Restricting our irregular state and holomorphic anomaly based results to 
$\epsilon_1=-\epsilon_2$ we have shown that Painlev\'{e} \RNum{1} 
$\tau$-function is related to partition function of ${\cal H}_0$ AD theory thus making above picture
fairly complete.

The paper is organized as follows:    
In section \ref{irr_states} we first review integer rank irregular states and then generalize 
this notion for the non-integer case of rank $5/2$ necessary to construct the dual 
2d CFT counterpart of ${\cal H}_0$ AD theory. We compute the corresponding irregular block 
and check that in the limit of vanishing $\Omega$-background it recovers the result 
obtained using SW curve approach.  

In section \ref{hol_recursion} we find $\Omega$-background corrections to the 
${\cal H}_0$ prepotential using holomorphic recursion method. We obtain exact in coupling 
expressions up to order $8$ in $\Omega$-background parameters. The results are in 
complete agreement with irregular state computations. 

In section \ref{NS_WKB} we discuss the important case of Nekrasov-Shatashvili limit 
using quasi-classical WKB method. 

The section \ref{painleve1}  is devoted to the remarkable relation between 
$\tau$-function of the 2nd order ODE Painlev\'{e} \RNum{1} and partition function of  ${\cal H}_0$ theory 
in restricted $\Omega$-background with $\epsilon_1=-\epsilon_2$.
%%%%%%%%%%%%%%%%%%%%%%%%%%%%%%%%%%%%%%%%%%%%%%%%%%%%%%%%%%%%%%%%%%%%%%%%%%%%%%%%%%%%%%%%%%%%%%%%%%
\section{Irregular conformal blocks}
\label{irr_states}
%%%%%%%%%%%%%%%%%%%%%%%%%%%%%%%%%%%%%%%%%%%%%%%%%%%%%%%%%%%%%%%%%%%%%%%%%%%%%%%%%%%%%%%%%%%%%%%%%%
\subsection{Irregular states}
The rank $n$ irregular states  $ |I^{(n)}\rangle $  in 2d Liouville conformal field theory, which 
depend on two sets of parameters ${\bf c}=\{c_0,\ldots ,c_n\}$ and ${\bm \beta}
=\{\beta_0, \ldots,\beta_{n-1}\}$, are defined by
\cite{Gaiotto:2012sf}
\bea
L_k |I^{(n)}( {\bf c},{\bm \beta} )\rangle  &=& {\cal L}^{(n)}_k|I^{(n)}( {\bf c},{\bm \beta} )\rangle \qquad ,\qquad k=0,\ldots n-1 \nn\\
L_k |I^{(n)}( {\bf c},{\bm \beta} )\rangle  &=& \Lambda^{(n)}_k|I^{(n)}( {\bf c},{\bm \beta} )\rangle \qquad ,\qquad k=n,\ldots 2n\nn\\
L_k |I^{(n)}( {\bf c},{\bm \beta} )\rangle  &=& 0  \qquad ,\qquad\qquad\qquad ~~~~k> 2n
\label{lneq}
\eea
with
\bea
{\cal L}^{(n)}_k & =&  (k+1)Qc_k -\sum_{\ell=0}^{k} c_\ell \, c_{k-\ell}
+\sum_{\ell=1}^{n-k}\ell c_{k+\ell}{\partial \over \partial c_{\ell} }  \qquad , \qquad k=0,\ldots , n-1 \nn\\
\Lambda^{(n)}_k   &=& (n+1)Qc_n\delta_{k,n} -\sum_{\ell=k-n}^{n}c_{\ell}  c_{k-\ell}  
\qquad \qquad \qquad ,\qquad  k=n,\ldots , 2n \label{lambdak}
\eea
As usual the parameter $Q$ is related to the central charge of Virasoro algebra by 
\[
c=1+6Q^2
\]
Notice that the differential operators ${\cal L}^{(n)}_k$ are constructed in 
such a way, that above relations are compatible with Virasoro algebra commutation 
rules. Namely, the form of this operators closely resembles the famous Feigin-Fuchs 
representation of Virasoro algebra in terms of free boson oscillators $c_k$. The meaning 
of the second set of parameters $\bm \beta $ is more subtle. These are remnants of internal 
Liouville momenta specifying successive OPE structure of those $n$ primary fields, 
which in colliding limit create the irregular state under discussion (see \cite{Gaiotto:2012sf} 
for details ).    

The state  $|I^{(n)}({\bf c},{\bm \beta} )\rangle $ can be expanded in $c_n$ power series
\be
|I^{(n)}( {\bf c},{\bm \beta} )\rangle  =
f({\bf c}, \beta_{n-1} )  \sum_{k=0}^\infty c_n^k   |I^{(n-1)}_k( \tilde {\bf c},\tilde {\bm\beta} ) \rangle
\ee
where 
\be
\tilde{\bf c} =(\beta_{n-1}  ,c_1,\ldots c_{n-1} ) \qquad , \qquad   \tilde{\beta}=(\beta_0,\beta_1\ldots \beta_{n-2})
\ee
and $  |I^{(n-1)}_k( \tilde {\bf c},\tilde {\bm\beta} ) \rangle $ is a level $k$ generalized descendant of rank 
$n-1$ irregular state  $ |I^{(n-1)}( \tilde {\bf c},\tilde {\bm\beta} )\rangle $ obtained by acting with Virasoro generators and derivatives with respect to $c_1,\ldots c_{n-1}$. It is argued in \cite{Gaiotto:2012sf} that 
after specifying the prefactor $f({\bf c}, \beta_{n-1} )$ appropriately these descendants can be determined order by order uniquely imposing equations (\ref{lneq}).
%%%
%%%
\subsection{Irregular conformal blocks}
The irregular conformal blocks defined as
\be
{\cal F}= \langle \Delta  |I^{(n)}( {\bf c},{\bm \beta} )\rangle.
\ee
will be related to the partition function of respective AGT-dual 4d gauge theory.
To identify this gauge theory following \cite{Alday:2009aq} one computes the (normalized) 
expectation value of the Liouville stress-energy tensor
\be
\phi_2(z)   =- { \langle  \Delta| T(z) | I^{(n)} \rangle  \over \langle \Delta  | I^{(n)}\rangle }
\ee
and treats $\sqrt{\phi_2(z)}dz$ as the Seiberg-Witten differential. The rank $2$ and $3$ 
which correspond to ${\cal H}_2$ and ${\cal H}_1$ Argyres-Douglas theories respectively,
have been investigated intensively in \cite{Gaiotto:2012sf,Nishinaka:2019nuy,Fucito:2023plp}. 
Surprisingly, the most basic case of Argyres-Douglas 
theory ${\cal H}_0$ is not addressed yet from this perspective. The reason is that in this case 
one deals with half-integer (namely  rank $5/2$ ) irregular states, but the corresponding 
representation theory is not developed yet. The main purpose of current work is to fill this gap. 
Though we'll mainly concentrate on rank $5/2$ case, our method of constructing irregular
states with half-integer rank seem to be quite general. 

\subsection{Irregular states of (Poincar\'{e}) rank 5/2}
Let us introduce a new type of irregular state, defined through relations 
\bea
\label{def_ir52}
L_k |I^{(5/2)}( c_1,c_2,\Lambda_5;\beta_0,c_0)\rangle  &=& 
{\cal L}_k|I^{(5/2)}(c_1,c_2,\Lambda_5;\beta_0,c_0)\rangle \qquad ,\qquad k=0,\ldots,5 \nn\\
L_k |I^{(5/2)}(c_1,c_2,\Lambda_5;\beta_0,c_0)\rangle  &=&
0  \qquad ,\qquad\qquad\qquad ~~~~k>5
\eea
with
\bea
{\cal L}_0 & =&c_1{\partial \over \partial c_{1} }
+2c_2{\partial \over \partial c_{2} }+5\Lambda_5{\partial \over \partial \Lambda_5 }\nn\\
{\cal L}_1 & =& \frac{2c_1^2c_2^2}{\Lambda_5}
+\frac{2c_2^3-3c_1\Lambda_5}{2c_2^2}{\partial \over \partial c_{1} }
+\frac{3\Lambda_5}{2c_2}{\partial \over \partial c_2 }\nn\\
{\cal L}_2 & =& \frac{\Lambda_5}{2c_2}{\partial \over \partial c_1 }\nn\\
{\cal L}_3 & =&-2c_1c_2\,;\quad {\cal L}_4 =-c_2^2\,;\quad {\cal L}_5 =-\Lambda_5
\label{ln_52}
\eea
Again these conditions are designed so that they are compatible with Virasoro algebra. 
It is natural to call them rank $5/2$ states since they exhibit behavior intermediate to 
rank 2 and rank 3 cases defined previously.
Similar to the integer rank cases we conjecture the following expansion for the rank $5/2$ irregular 
state to be hold
\be
\label{rank52_vs_rank2_expansion}
|I^{(5/2)}(c_1,c_2,\Lambda_5;\beta_0,c_0)\rangle  =
f(c_0,c_1,c_2,\Lambda_5 )  \sum_{k=0}^\infty \Lambda_5^k   |I^{(2)}_k(c_0,c_1,c_2;\beta_0,\beta_1 ) \rangle
\ee
The leading term $|I^{(2)}_0(c_0,c_1,c_2;\beta_0,\beta_1 ) \rangle$ is just the rank 2 irregular state, while 
the generalized descendants are some linear combinations of monomials \footnote{Given a partition 
$Y=1^{n_1}2^{n_2}3^{n_3}\cdots$, by definition $L_{-Y}=\cdots L_{-3}^{n_3}L_{-2}^{n_2}L_{-1}^{n_1}$.}
\bea 
L_{-Y}c_1^{r_1}c_2^{-r_2}\partial_{c_1}^{m_1}\partial_{c_2}^{m_2}|I^{(2)}_0(c_0,c_1,c_2;\beta_0,\beta_1 ) \rangle
\eea  
where $n=|Y|$, $r_{1,2}$,$m_{1,2}$ are non-negative integers, subject to constraint 
\bea 
5k=n+m_1+2m_2+2r_2-r_1
\eea
In addition, the maximal power of $c_1 $ for given level $k$ is restricted by $r_1\le 3k$. 
Thus, though the number of allowed terms grows drastically with the level, still for a given $k$ 
it is finite. In fact the parameter $\beta_1$ does not show up itself in expansion
(\ref{rank52_vs_rank2_expansion}), so from now on we'll omit it in arguments of irregular states.
   
Finding the appropriate prefactor $f(c_0,c_1,c_2,\Lambda_5 )$, which summarizes all 
non-analytical in $\Lambda_5$ part, was a challenging task. To identify this factor in particular 
we have carefully analyzed the small $\Lambda_5$ behavior of corresponding gauge theory prepotential using 
Seiberg-Witten curve method (see discussion at the end of section \ref{H0_SW_exact}). Here is the final outcome: 
in order to be consistent with (\ref{def_ir52}), (\ref{ln_52}), the prefactor should be chosen as
\be
\label{factor_f}
f(c_0,c_1,c_2,\Lambda_5 )=c_2^{\rho_2}\Lambda_5^{\rho_5}\exp (S(c_0,c_1,c_2,\Lambda_5))
\ee
with
\bea
\label{function_S} 
S(c_0,c_1,c_2,\Lambda_5)&=&-\frac{2 c_1^2 c_2^4}{3 \Lambda _5^2}+\frac{4 c_1 c_2^7}{27 \Lambda _5^3}
+\frac{4 \left(c_2^4-6 c_1 c_2 \Lambda _5\right){}^{5/2}-4 c_2^{10}}{405 \Lambda _5^4}\nn\\
&+&\frac{8 \left(c_2^4-6 c_1 c_2 \Lambda _5\right){}^{5/4} \left(c_0
-\frac{3 Q}{2}\right)}{15 \Lambda _5^2}+\frac{c_1^2 \left(Q-c_0\right)}{c_2}
\eea 
Acting by the Virasoro zero mode $L_0$ and comparing the left and right sides of expansion
(\ref{rank52_vs_rank2_expansion}) we see that the constants $\rho_2$, $\rho_5$ are related 
by 
\bea
\label{rho25} 
2\rho_2+5\rho_5=c_0(Q-c_0)
\eea
Acting by the operators $L_1$ and $L_2$ 
using (\ref{def_ir52}), (\ref{ln_52}) on left and 
(\ref{lneq}) with $n=2$, on  right sides, we get recursion relation, connecting
the level $k$ descendant $|I^{(2)}_k \rangle $ 
with lower level descendants. Though we do not have a rigorous proof, 
through extensive calculations up to level 5, we get convinced that these relations 
are strong enough to determine the descendants uniquely    
much like in the cases with integer rank irregular states discussed in 
\cite{Gaiotto:2012sf,Nishinaka:2019nuy}.
We have observed that for odd $k$, knowing $|I^{(2)}_{k-1} \rangle $ and using  
recursion we get $|I^{(2)}_k \rangle $ uniquely. Instead, level $k$ 
calculation with $k$ even leaves one coefficient, namely the one in front of the term $c_2^{-5k/2}$
undefined. This coefficient gets uniquely determined at the next level.
Besides, already at level 2, for the parameter $\rho_2$ we find    
\bea
\label{rho2}
\rho_2=c_0(2 c_0-7Q)+\frac{71}{12}\, Q^2-\frac{1}{12}
\eea
Here is the result for level one descendant
\bea 
\label{desc_1}
&&|I^{(2)}_1(c_0,c_1,c_2;\beta_0 ) \rangle =\left[\frac{1}{6 c_2^2}\,L_{-1}-\frac{5 c_1}{6 c_2^2}\partial_{c_2}
+\left(\frac{c_1^2}{2 c_2^3}-\frac{2( c_0-3Q/2)}{3 c_2^2}\right)\partial_{c_1}\right.\nn\\
&&\qquad \left.-\frac{c_1 \left(-60 c_0 Q+16 c_0^2+55 Q^2-1\right)}{8 c_2^3}-\frac{11 c_1^3 \left(c_0-Q\right)}{6 c_2^4}\right]|I^{(2)}(c_0,c_1,c_2;\beta_0 )\rangle \qquad\,.
\eea
Explicit forms of level $2$ and $3$ descendants are given in the appendix \ref{rank2_desc}.
%%%%%%%%%%%%%%%%%%%%%%%%%%%%%%%%%%%%%%%%%%%%%%%%%%%%%%%%%%%%%%%%%%%%%%%%%%%%%%%%%%%%%%%%%%%%%%%%%%%%%%%%%
\subsection{ ${\cal H}_0$ AD theory}
%%%%%%%%%%%%%%%%%%%%%%%%%%%%%%%%%%%%%%%%%%%%%%%%%%%%%%%%%%%%%%%%%%%%%%%%%%%%%%%%%%%%%%%%%%%%%%%%%%%%%%%%%
The conformal block, which will be related to the partition function of the ${\cal H}_0$ 
AD theory is defined as\footnote{In order to get a non-vanishing result after pairing 
with the vacuum state $\langle 0|$ one should set the Liouville charge parameter $\beta_0=0$.}
\be
\label{ZH0_def}
{\cal Z}_{{\cal H}_0} =\langle  0  |I^{(5/2)}(c_1,c_2,\Lambda_5;0,c_0)\rangle
\ee
To proceed we need to calculate the vacuum amplitude $\langle 0 |I^{(2)}\rangle $.
The strategy is to insert generators $L_{0,1}$ which 
annihilate the left vacuum while on the right act by the differential operators 
${\cal L}^{(2)}_{0,1}$ defined in (\ref{lambdak}). We get two differential relations
\bea
&&c_0(Q-c_0)+\left(c_1\partial_{c_1}+2c_2\partial_{c_2}\right)\log \langle 0 |I^{(2)}\rangle =0\nn\\
&&2c_1(Q-c_0)+c_2\partial_{c_1}\log \langle 0 |I^{(2)}\rangle =0
\eea    
which up to an inessential $c_{1,2}$ independent constant multiplier give 
\bea 
\label{vac_ir2}
\langle 0 |I^{(2)}\rangle =c_2^{-\frac{c_0(Q-c_0)}{2}} e^{-\frac{c_1^2 (Q-c_0)}{c_2}}
\eea  
Plugging (\ref{rank52_vs_rank2_expansion}) into (\ref{ZH0_def}) and taking into account 
(\ref{factor_f}), (\ref{desc_1}) and, (\ref{vac_ir2}) one finds $Z_{{\cal H}_0}=Z_{{\cal H}_0,\rm tree}Z_{{\cal H}_0,\rm inst}$ with
\bea
\label{Zh0tree}
Z_{ {\cal H}_0 \rm tree} &=& c_2^{-\frac{c_0(Q-c_0)}{2}+\rho_2}\Lambda_5^{\rho_5}
e^{-\frac{c_1^2 (Q-c_0)}{c_2}+S}\\
\label{Zh0inst}
Z_{ {\cal H}_0 \rm inst} &=&
1{+}\frac{c_1}{8 c_2^3}\,\left(1-71Q^2+30 c_0 (3Q-c_0)\right) \Lambda _5 {+}\ldots
\eea
and $S$, $\rho_2$, $\rho_3$ given in  (\ref{function_S}), (\ref{rho25}), (\ref{rho2}).
For the sake of simplicity the higher order in $\Lambda_5$ are omitted here. Since this terms 
are needed for comparison with results obtained from holomorphic anomaly or from Painlev\'{e} I
$\tau$-function, we display few of them in appendix \ref{irr_block_4} explicitly.
 
For the normalized expectation value of the stress tensor, which defines the SW-differential 
of gauge theory we have
\bea
\phi_2(z) &=&- { \langle 0 | T(z) |I^{(5/2)}\rangle  \over  \langle 0 ||I^{(5/2)}\rangle } =
{ 2v \over z^4}  {+}{ 2 c_1 c_2 \over z^5}
{+}{c_2^2\over z^6}
{+}{\Lambda_5 \over z^7} \label{phi2h0}
\eea
with
\be
\label{G_Matone_CFT}
v=- \frac{\Lambda_5}{4c_2}\,\partial_{c_1}\log{\cal Z}_{{\cal H}_0} 
\ee

Let us first perform the simplest check against Seiberg-Witten curve analysis.  
In a usual way we introduce gauge theory like parameters as
\be
\label{map_CFT_H0} 
Q=\frac{s}{\sqrt{p}};~ c_0=\frac{a+{3s\over 2}}{\sqrt{p}};
~v ={\hat{v} \over p} ; ~\phi_2=\frac{\hat\phi_2}{p}; ~\Lambda_5=\frac{\hat\Lambda_5}{p}; 
~  c_i=\frac{\hat{c}_i}{\sqrt{p}};~i=1,2
\ee
where $s=\epsilon_1+\epsilon_2$ and  $p=\epsilon_1\epsilon_2$. Then we have
\bea
\hat\phi_2(z) &=&
{ 2\hat{v} \over z^4}  {+}{ 2 \hat{c}_1 \hat{c}_2 \over z^5}
{+}{\hat{c}_2^2 \over z^6}
{+}{ \hat{\Lambda}_5\over z^7} \label{phi2hat}
\eea
The $1$-form
\be 
\label{SW_dif}
\lambda_{SW}=\sqrt{\hat\phi_2(z)}\,dz
\ee 
is the Seiberg-Witten differential. The period integrals along $A$ and 
$B$-cycles can be evaluated exactly in terms of hypergeometric 
function (see section \ref{H0_SW_exact}), but for the present purposes it is sufficient to notice, that $A$-cycle shrinks to 
the point $z=0$ in $\Lambda_5\to 0$ limit, so that in this case one can simply expand
$\sqrt{\hat\phi_2}$ in powers of $ \hat{\Lambda}_5$ and then take the residues at $z=0$. 
Here is the result up to order $O(\hat{\Lambda}_5^2)$:
\be
a ={1\over 2\pi i} \oint_{z=0} \sqrt{\hat\phi_2} dz  =
\frac{\hat{v}}{\hat{c}_2}-\frac{\hat{c}_1^2}{2\hat{c}_2}+
\left(\frac{3 \hat{c}_1 \hat{v}}{2\hat{c}_2^4}-\frac{5\hat{c}_1^3}{4\hat{c}_2^4}\right) \hat{\Lambda}_5 +\ldots
 \ee
Inverting for $\hat{v}$ one finds
\be
\hat{v}=a \hat{c}_2+\frac{\hat{c}_1^2}{2}+\left(\frac{\hat{c}_1^3}{2 \hat{c}_2^3}
-\frac{3 a \hat{c}_1}{2 \hat{c}_2^2}\right)\hat{\Lambda}_5 +\ldots
\ee
This nicely matches the result for $v$ obtained by plugging (\ref{Zh0tree}),  
(\ref{Zh0inst}) into (\ref{G_Matone_CFT})
\be
v=c_0 c_2+\frac{c_1^2}{2}+\left(\frac{c_1^3}{2 c_2^3}
-\frac{3c_0 c_1}{2 c_2^2}\right)\Lambda _5+\ldots
\ee
 taking into account (\ref{map_CFT_H0}). In the forthcoming sections we will see that this agreement holds also in presence of $\epsilon$-corrections.
%%%%%%%%%%%%%%%%%%%%%%%%%%%%%%%%%%%%%%%%%%%%%%%%%%%%%%%%%%%%%%%%%%%%%%%%%%%%%%%%%%%%%%%%%%%%%%%%%%%%%%%%%
\section{ The holomorphic anomaly recursion }
\label{hol_recursion}
%%%%%%%%%%%%%%%%%%%%%%%%%%%%%%%%%%%%%%%%%%%%%%%%%%%%%%%%%%%%%%%%%%%%%%%%%%%%%%%%%%%%%%%%%%%%%%%%%%%%%%%%%
In this section we derive formula for the prepotential of ${\cal H}_0$ theory which is exact in coupling 
but dependence on $\Omega$-background is given order by order as power series 
in $\epsilon_{1,2}$. Notice that CFT approach discussed in previous section provides 
a complementary framework: we have power series in coupling with coefficients, exact 
in $\epsilon_{1,2}$. 

We will closely follow the presentation in \cite{Fucito:2023plp} 
where, instead AD theories ${\cal H}_{1,2}$ were investigated.

For the full prepotential we have
\bea
{\cal F}=\epsilon_1\epsilon_2 \log Z=\sum_{n=0,m=0}^\infty \left(\epsilon_1+\epsilon_2\right)^{2n}
\left(\epsilon_1\epsilon_2\right)^{m}F^{(n,m)} =\sum_{g=0}^{\infty}
\left(\epsilon_1\epsilon_2\right)^{g}{\cal F}_g
\eea
where
\be
{\cal F}_g=\sum_{n+m=g}\left({s^{2}\over p}\right)^nF^{(n,m)}
\ee  
and we parameterize  the $\Omega$-background with variables
\be
s=\epsilon_1+\epsilon_2 \qquad, \qquad p=\epsilon_1 \epsilon_2
\ee

\subsection{The SW prepotential ${\cal F}_0$}

The term ${\cal F}_0$ which does not  depend on $\epsilon_{1,2}$ is just the SW prepotential. 
As it is shown in section \ref{H0_SW_exact}, the SW differential (\ref{phi2hat}), (\ref{SW_dif}) can be 
cast into canonical form (\ref{SW_dif_canonical}). The periods $a(\hat{v})$ and  
$a_D(\hat{v})$ are expressed in terms of Gauss hypergeometric functions 
(see (\ref{a}), (\ref{aD}) ). Then ${\cal F}_0$ can be found using the relations  
(\ref{a_D_v_vs_F}).   

Let us keep discussion in this section more general and consider any SW theory governed 
by an elliptic curve. Suppose this elliptic curve is cast in Weierstrass 
canonical form
\be
y^2 =4 z^3 - g_2 z- g_3
\ee
where $g_2$ and $g_3$ are polynomials in global modulus parameter $u$\footnote{
In case of AD theory ${\cal H}_0$  the role of $u$ is played by $\hat{v}$, $g_2$ 
is $\hat{v}$ independent and $g_3$ is a linear in $\hat{v}$ (see 
(\ref{g2_g3_H0})). }.
Periods of the Weierstrass elliptic curve
are given by
\be
\omega_i =\oint_{\gamma_i} dz/(i \pi y) 
\ee 
where $\gamma_1$ and $\gamma_2$ are $A$ and $B$ cycles of the torus\footnote{In ${\cal H}_0$ case
$\omega_1=\partial_{\hat{v}}a$, $\omega_2=\partial_{\hat{v}}a_D$ (see (\ref{dua}), (\ref{duaD})).}.
As usual the infrared coupling $\tau_{IR}$ is identified with torus parameter 
$\tau_{IR}= {\omega_2\over \omega_1}$. It is convenient to introduce 
the nome given by $q=e^{\pi {\rm i}\tau_{IR}}$.  Due to standard formulae 
of elliptic geometry\footnote {The Eisenstein series are given by $E_k(q)=
1+{2\over \zeta (1-k)}\sum _{n=1}^{\infty} \frac{n^{k-1} q^{2 n}}{1-q^{2 n}}$
\,, $k=2,4,6,\cdots $\,.}
\bea
g_2=\frac{4}{3\omega_1^4}E_4(q)\,;\qquad 
g_3=\frac{8}{27\omega_1^6}E_6(q)
\eea 
In particular we have important relations 
\be
{27 g_3^2\over g_2^3} = { E_6(q)^2\over E_4(q)^3} \qquad , 
\qquad \omega_1(q,u)^2= \frac{2g_2E_6(q)} {9 g_3E_4(q)} \label{w1ee}
\ee
 In particular from the first equation one finds
 \be
 \label{qdqu}
D_\tau u\equiv q \partial_q u  = {2\left(E_4^3-E_6^2\right) 
\over E_4 E_6 \left( {3g_2'(u) \over g_2(u)} -
 {2g_3'(u)  \over g_3(u) } \right)}
\ee
where we have used Ramanujan differentiation rules
\be
\label{qdqE46} 
\quad D_\tau E_4 = \ft{2}{3} (E_2 \, E_4-E_6)  \quad , \quad
D_\tau E_6=  E_2 \, E_6-E_4^2\,.
\ee
For later purposes let us remind also differentiation rule for the degree $2$ quasi-modular 
form
\be
\label{qdqE2}  
D_\tau E_2 = \ft{1}{6} (E_2^2-E_4)
\ee
The "flat" coordinate $a$ and the SW prepotential ${\cal F}(a)$ are introduced through standard relations 
\be
{\cal F}''(a)=-2\log q \qquad , \qquad    \omega_1(q,u)={da\over du}
\ee

 \subsection{ ${\cal F}_g$-terms}

 Higher order terms can be computed recursively using holomorphic anomaly relation
 \be\label{HAE}
 \partial_{E_2} {\cal F}_{g}  =\ft{1}{24} \left[    \partial_a^2 {\cal F}_{g-1} +\sum_{g'=1}^{g-1} \partial_a {\cal F}_{g'} \partial_a {\cal F}_{g-g'}  \right]
\ee
starting from $g=1$ expression
\be
{\cal F}_1 (u,b,q)={1 \over 4}  \log {1\over\omega_1(q,u)^2} 
+\ft{ s^2-2p }{24p}\log \Delta (u) \label{f1h1}
\ee
where
\be
\Delta(u)=g_2^3
-27g_3^2
\ee 
is the modular discriminant.

Following \cite{Huang:2006si,Huang:2009md,Huang:2011qx,Huang:2013eja} we introduce the quantities
\be
S={2\over 9\omega_1(q,u)^2 }=\frac{g_3(u)E_4(q)}{g_2(u)E_6(q)}  \qquad ,\qquad  X = S  E_2(q)
\ee
Their total $u$-derivatives can be computed using the equations (\ref{qdqu}), 
(\ref{qdqE46}), (\ref{qdqE2}). Here are the results:
\bea
p_1 &=& {d\over du} \ln S =\frac{9 X \left(2 g_2 g_3'-3 g_3 g_2'\right)+g_2^2 g_2'-18 g_3 g_3'}{2 \left(g_2^3-27
   g_3^2\right)} \\
p_2 &=&{dX\over du}  =\frac{27 X^2 \left(2 g_2 g_3'-3 g_3 g_2'\right)+6 X \left(g_2^2 g_2'-18 g_3
   g_3'\right)+g_2 \left(2 g_2 g_3'-3 g_3 g_2'\right)}{12 \left(g_2^3-27 g_3^2\right)}\nn
\eea
It is easy to check that the derivatives of $a$ with respect to $u$, in terms of quantities 
introduced above,  are given by
\bea
\left(\frac{du}{da}\right)^2 &=& {1 \over  \omega_1^{2} } 
=\frac{9S}{2}\qquad , \qquad \frac{d^2u}{da^2} =\frac{1}{2}\frac{d}{d u}\, \omega_1^{-2}
=\frac{9S\,p_1}{4}  \label{d12au}
\eea
This allows to rewrite (\ref{HAE}) in a more convenient form
\bea\label{HAE2}
	\partial_{X} {\cal F}_{g}
	&=&  \frac{3}{16} \left( D_u^2  {\cal F}_{g-1}+ {p_1\over 2} D_u  {\cal F}_{g-1}
	+  \sum_{g'=1}^{g-1}  D_u  {\cal F}_{g'}D_u  {\cal F}_{g-g'} \right)\,.
	\eea
	In this setting one should consider ${\cal F}_{g}$ as functions of two independent variables 
	$u$ and $X$.
	The total derivative  $D_u$ is
	\bea 
	D_u=\frac{\partial}{\partial u}+p_2\frac{\partial}{\partial_X}\,.
\eea 	
In this setting one should consider ${\cal F}_{g}$ as functions of two independent variables 
$u$ and $X$. A careful analysis carried out in \cite{Huang:2011qx} shows that
${\cal F}_{g}$ is a polynomial in $X$ of maximal degree $3(g-1)$ with rational in $u$ 
coefficients. More precisely the denominators of this coefficients are equal 
to $\Delta(u)^{2g-2}$ and numerators are polynomials in $u$ of maximal degree 
$2d_{\Delta}(g-1)-1$, where $d_{\Delta}$ is the degree of discriminant in $u$.
Evidently, the equation (\ref{HAE2}) alone can not fix $X$ independent terms. This ambiguity 
can be removed imposing so called gap condition. Namely, for $g>1$ near each zero $u_*$ 
of the discriminant, the gap conditions reads
\bea\label{gap_condition}
{\cal F}_g  & \underset{ u\to  u^*} {\approx}  &  {\frac{(2g-3)!}{a^{2g-2}}  \sum_{k=0}^{g} \hat B_{2k} \hat B_{2g-2k}\left(\frac{\epsilon_1 }{\epsilon_2}\right)^{g-2k}}  + O( a^0)
\eea
where
\be
\hat{B}_m=\left( 2^{1-m}-1\right) {B_m\over m!}
\ee
with $B_m$ the Bernoulli numbers and $a$ is the local flat coordinate, vanishing 
at $u=u^*$. 

Notice the absence of lower order poles 
$a^{-n}$ with $n<2g-2$ in (\ref{gap_condition}), hence the term "gap condition". 
In the next section, using above described scheme we will find explicit 
expressions for ${\cal F}_{1,2,3}$. We'll also check that
they agree with the result obtained from the irregular state approach.

 %%%%%%%%%%%%%%%%%%%%%%%%%%%%%%%%%%%%%%%%%%%%%%%%%%%%%%%%%%%%%%%%%%%%%%%%%%%%%%%%%%%%%%%%%%%%%%%%
\subsection{Holomorphic anomaly recursion for ${\cal H}_0$ theory}
Here we apply the method described in previous section for the case of 
our main interest ${\cal H}_0$ theory.  
%%%%%%%%%%%%%%%%%%%%%%%%%%%%%%%%%%%%%%%%%%%%%%%%%%%%%%%%%%%%%%%%%%%%%%%%%%%%%%%%%%%%%%%%%%%%%%%%
\subsection{${\cal H}_0$ in flat background}
\label{H0_SW_exact}
Our starting point is the Seiberg-Witten differential
\be 
\lambda_{SW}=\sqrt{\hat{\phi}}_2 \,{dz\over 2\pi i}
\ee
with $\hat{\phi}_2$ given in (\ref{phi2hat}). To bring the curve into canonical form 
let us perform change of variable
\be 
\label{SW_dif_canonical}
z=-\frac{3\hat{\Lambda}_5}{3x+\hat{c}_2^2}
\ee
Then for SW differential we get
\be 
\lambda_{SW}=\frac{1}{2\hat{\Lambda}_5^{2}}\sqrt{-4x^3+g_2x+g_3}\,{dx\over 2\pi i}
\ee
with Weierstrass parameters
\be 
\label{g2_g3_H0}
g_2=\frac{4 \hat{c}_2^4}{3}-8 \hat{c}_1 \hat{c}_2 \hat{\Lambda}_5\,; \quad 
g_3=-\frac{8}{3} \hat{c}_1 \hat{c}_2^3 \hat{\Lambda}_5+\frac{8 \hat{c}_2^6}{27}+8 \hat{\Lambda}_5^2 \hat{v}
\ee 
Notice that in the limit $\hat{\Lambda}_5\to 0$ one can chose a small contour surrounding 
$x=-\hat{c}_2^2/3$ anticlockwise, as the $A$-cycle.

For the holomorphic differential we simply have
\be 
\partial_{\hat{v}}\lambda_{SW}=\frac{2}{\sqrt{-4x^3+g_2x+g_3}}\,{dx\over 2\pi i}
\ee
The periods of this holomorphic differential can be expressed in terms of 
the hypergeometric function
\bea 
\label{dua}
\partial_{\hat{v}}a&=&
\left(\frac{3 g_2}{4}\right){}^{-\frac{1}{4}}\,
_2F_1\left(\frac{1}{6},\frac{5}{6};1;\frac{1}{2}-\frac{1}{2}\sqrt{\frac{27 g_3^2}{g_2^3}}\right)\\
\label{duaD}
\partial_{\hat{v}}a_D&=&i \left(\frac{3 g_2}{4}\right){}^{-\frac{1}{4}} \,
_2F_1\left(\frac{1}{6},\frac{5}{6};1;\frac{1}{2}+\frac{1}{2}\sqrt{\frac{27 g_3^2}{g_2^3}}\right)
\eea 
Remarkably above expressions can be easily integrated over $\hat{v}$ to get periods of $\lambda_{SW}$
\footnote{To check that the integration constants are chosen correctly one can e.g. consider the 
	limit $\hat{\Lambda}_5\to 0 $. }
\bea
\label{a} 
a&=&-\frac{1}{27 \hat{\Lambda}_5^2}
\left(\frac{3 g_2}{4}\right)^{{5\over 4}}\left(1-\sqrt{\frac{27 g_3^2}{g_2^3}}\right) \,
_2F_1\left(\frac{1}{6},\frac{5}{6};2;\frac{1}{2}-\frac{1}{2}\sqrt{\frac{27 g_3^2}{g_2^3}}\right)\\
\label{aD}
a_D&=&\frac{i}{27 \hat{\Lambda}_5^2}
\left(\frac{3 g_2}{4}\right)^{{5\over 4}}\left(1+\sqrt{\frac{27 g_3^2}{g_2^3}}\right) \,
_2F_1\left(\frac{1}{6},\frac{5}{6};2;\frac{1}{2}+\frac{1}{2}\sqrt{\frac{27 g_3^2}{g_2^3}}\right)
\eea
The formulae (\ref{dua}) a (\ref{a}) are well suited to perform small $\hat{\Lambda}_5$ expansion 
(in this limit the argument of hypergeometric function approaches to zero). Instead for 
dual periods  (\ref{duaD}) and (\ref{aD}) it is convenient to use the formulae
\bea 
&&\Gamma (v) \Gamma (1-v)_2F_1\left(v,1-v;1;1-x\right)=-\log (x)\, _2F_1(v,1-v;1;x)\nn\\
&&\qquad-\sum _{n=0}^{\infty}\frac{(v)_n (1-v)_n}{(n!)^2}
\left(\psi(1+n-v)+\psi(n+v)-2\psi(1+n)\right)\, x^n\\
&&\Gamma (1+v) \Gamma (2-v)_2F_1\left(v,1-v;2;1-x\right)=v(1-v)x\log (x)\, _2F_1(1+v,2-v;2;x)\nn\\
&&\qquad+1+\sum _{n=1}^{\infty}\frac{(v)_n (1-v)_n}{n!(n-1)!}
\left(\psi(1+n-v)+\psi(n+v)-\psi(1+n)-\psi(n)\right)\, x^n\hspace{3cm}
\eea
where
\be 
\psi(x)=\frac{d}{dx}\log \Gamma(x)
\ee 
Using above formulae we have checked that  
\bea 
\label{closed_1_form}
\partial_{\hat{v}}a \,\partial_{\hat{c}_1}a_D
-\partial_{\hat{v}}a_D \,\partial_{\hat{c}_1}a=-\frac{2i\hat{c}_2}{\pi\hat{\Lambda}_5}
\eea
This equation can be rewritten as
\bea 
d\left( a_D da -\frac{2i\hat{c}_2}{\pi \hat{\Lambda}_5}\,\hat{v}d\hat{c}_1 \right)=0
\eea  
where $a$ and $a_D$ are considered as functions on two-dimensional manifold 
with coordinates $(\hat{v},\hat{c}_1)$ and $d$ is the external differential. Since the $1$-form 
in brackets is closed it can be represented (locally) as differential of 
some function ${\cal F}$ 
\bea 
a_Dda-\frac{2i\hat{c}_2}{\pi \hat{\Lambda}_5}\,\hat{v}d\hat{c}_1 =d \left(\frac{i}{2\pi }\, {\cal F}\right)
\eea
If considered as a function of $(a,\hat{c}_1)$, instead of $\hat{v},\hat{c}_1$, from above equation 
we have
\bea 
\label{a_D_v_vs_F}
a_D=\frac{i}{2\pi }\,\partial_{a} {\cal F}_0\,;\qquad \hat{v}
=-\frac{\hat{\Lambda}_5}{4\hat{c}_2}\partial_{\hat{c}_1} {\cal F}_0\,.
\eea 
The first equality shows that  ${\cal F}_0$ is just the prepotential, while the second 
equality coincides with relation (\ref{G_Matone_CFT}).

Let us conclude this section with an observation that helped us to identify
the function $S(c_0,c_1,c_2,\Lambda_5)$ in (\ref{function_S}). This was an important step 
in constructing the rank $5/2$ irregular state. Analyzing $\hat{\Lambda}_5\to 0 $ limit 
of (\ref{a}) we see that the argument of hypergeometric function approaches to zero, hence 
substituting it by $1$, for the singular part we get
\bea 
a&\sim &-\frac{1}{27 \hat{\Lambda}_5^2}
\left(\frac{3 g_2}{4}\right)^{{5\over 4}}\left(1-\sqrt{\frac{27 g_3^2}{g_2^3}}\right)
\eea 
which can be easily inverted with result
\bea 
\hat{v}\sim \frac{\left(\hat{c}_2^4-6 \hat{c}_1 \hat{c}_2 \Lambda _5\right)^{3/2}}{27 \Lambda _5^2}
-\frac{\hat{c}_2^3 \left(\hat{c}_2^3-9 \hat{c}_1 \hat{\Lambda}_5\right)}{27 \hat{\Lambda}_5^2}
-a \left(\hat{c}_2^4-6 \hat{c}_1 \hat{c}_2 \Lambda _5\right)^{1/4}
\eea 
From the second equality in (\ref{a_D_v_vs_F}) for the non-analytic part of 
prepotential we find
\bea 
{\cal F}&=&\int -\frac{4 \hat{c}_2}{\hat{\Lambda}_5}\hat{v}d\hat{c}_1\sim \nn\\
&-&\frac{8 a \left(\hat{c}_2^4-6 \hat{c}_1 \hat{c}_2\hat{\Lambda}_5\right)^{5/4}}{15\hat{\Lambda}_5^2}
+\frac{4 \left(\hat{c}_2^4-6 \hat{c}_1 \hat{c}_2\hat{\Lambda}_5\right){}^{5/2}}{405\hat{\Lambda}_5^4}
-\frac{2 \hat{c}_1^2 \hat{c}_2^4}{3\hat{\Lambda}_5^2}+\frac{4 \hat{c}_1 \hat{c}_2^7}{27\hat{\Lambda}_5^3}
-\frac{4 \hat{c}_2^{10}}{405\hat{\Lambda}_5^4}
\eea 
where the last $\hat{c}_1$ independent term is added to cancel dangerous forth order 
pole in $\hat{\Lambda}_5$. Thus in view of the map (\ref{map_CFT_H0}), we have 
recovered (\ref{function_S}).  
%%%%%%%%%%%%%%%%%%%%%%%%%%%%%%%%%%%%%%%%%%%%%%%%%%%%%%%%%%%%%%%%%%%%%%%%%%%%%%%%%%%%%%%
\subsection{Corrections in $\epsilon_{1,2}$}
In this section we  derive $q$-exact formulae for the first few ${\cal F}_g$-terms 
using  the holomorphic recursive algorithm. The results
will be checked against those obtained in the previous section using irregular state approach.
The dynamics of ${\cal H}_0$ theory is governed by the Weierstrass elliptic curve 
\be
y^2 =4 x^3 -g_2 x-g_3
\ee
with parameters (\ref{g2_g3_H0}). We see that $g_2$ is $\hat{v}$-independent and 
$g_3$ is linear in $\hat{v}$. Consequently the discriminant has two simple 
zeros. Applying method described in previous section we get
\begin{scriptsize}
\bea
\label{prep_exact}
&&-\ft12{\partial^2 {\cal F}_0 \over \partial a^2} = \log q\nn\\
&&{\cal F}_1=\frac{s^2-2 p}{24p} \log \left(g_2^3-27 g_3^2\right)
+\frac{1}{4} \log \frac{9g_3E_4}{2g_2E_6}\nn\\
&&{\cal F}_2=\frac{\Lambda _5^4g_2^2g_3}{\left(g_2^3-27 g_3^2\right){}^2}  
\left(\frac{15E_2^3 }{4E_6}+\frac{9\left(11 p-2 s^2\right) E_2^2}{4pE_4} 
\right.\nn\\
&&\left.+\frac{9\left(11 p^2-12 p s^2+s^4\right) E_6 E_2}{4p^2E_4^2}+
\frac{9\left(7 p-6 s^2\right)E_4 E_2}{2pE_6} 
+\frac{3}{20p^2} \left(299 p^2-618 p s^2+237 s^4\right)\nn
\right)\nn\\
&&{\cal F}_3=\frac{16 \Lambda _5^8g_2^7}{9\left(g_2^3-27 g_3^2\right){}^4} 
\left(\frac{135E_2^6}{64 E_4^3}+
\frac{135\left(16 p-s^2\right) E_2^5 E_6}{64pE_4^4}
+\frac{81\left(28 p-5 s^2\right)E_2^4}{32pE_4^2}\right.\nn\\
&&\left.+\frac{27\left(477 p^2-77 p s^2+2 s^4\right)E_6^2E_2^4}{64p^2E_4^5}
+\frac{9 \left(-1325 p^2 s^2+3630 p^3+90 p s^4-s^6\right) E_6^3E_2^3 }{64p^3E_4^6} \right.\nn\\
&&+\frac{9\left(6242 p^2-2581 p s^2+72 s^4\right) E_6E_2^3}{64p^2E_4^3}
+\frac{27\left(8023 p^2-7596 p s^2+654 s^4\right)E_2^2}{320p^2E_4}\nn\\
&&+\frac{27\left(-39363 p^2 s^2+39964 p^3+3752 p s^4-30 s^6\right) E_6^2 E_2^2}{320p^3E_4^4}
+\frac{135\left(p-s^2\right) \left(11 p-s^2\right) \left(16 p-s^2\right) E_6^4 E_2^2}{64p^3E_4^7}
\nn\\
&&+\frac{27\left(-92881 p^2 s^2+44926 p^3+38066 p s^4-1427 s^6\right) 
E_6^3E_2}{320p^3E_4^5}+\frac{27\left(-47300 p^2 s^2+24273 p^3+17772 p s^4
-237 s^6\right) E_6 E_2}{160p^3E_4^2}\nn\\ \nn
&&
+\frac{9 \left(171350 p^3-564379 p^2 s^2+456678 p s^4-100998 s^6\right)}{2240 p^3}+\frac{9  \left(154373 p^3-519794 p^2 s^2+426750 p s^4-95462 s^6\right)E_6^2}{320 E_4^3 p^3}
\\ && \left.
 +
\frac{9 \left(34210 p^3-117270 p^2 s^2+97500 p s^4-21983 s^6\right) E_6^4}{320 E_4^6 p^3}
\right)
\eea 
\end{scriptsize}
In order to express ${\cal F}_g$ as a function of flat modulus $a$, one can find 
$q$ as function of $\hat{v}$ inserting (\ref{dua}), (\ref{duaD}) in
\bea 
q=\exp \left(\pi i \frac{\partial_{\hat{v}} a_D}{\partial_{\hat{v}} a}\right)
\eea 
and then inverting (\ref{a}) to express $\hat{v}$ in terms of $a$.

In the limit $\hat{\Lambda}_5 \to 0$ we get
\bea
q=\frac{\hat{\Lambda}_5 \sqrt{\hat{c}_1^2-2 \hat{v}}}{8 \hat{c}_2^3}
+\frac{\hat{\Lambda}_5^2 \left(5 \hat{c}_1^3
-9 \hat{c}_1 \hat{v}\right)}{8 \hat{c}_2^6 \sqrt{\hat{c}_1^2-2 \hat{v}}}
+\frac{\hat{\Lambda}_5^3 \left(1980 \hat{c}_1^2 \hat{v}^2-1854 \hat{c}_1^4\hat{v}
+469 \hat{c}_1^6-312 \hat{v}^3\right)}{128 \hat{c}_2^9 \left(\hat{c}_1^2-2 \hat{v}\right){}^{3/2}}\qquad\nn\\
+O(\hat{\Lambda}_5^4)\qquad
\eea 
and
\bea 
\hat{v}=a \hat{c}_2+\frac{\hat{c}_1^2}{2}+\hat{\Lambda}_5 \left(\frac{\hat{c}_1^3}{2 \hat{c}_2^3}
-\frac{3 a \hat{c}_1}{2 \hat{c}_2^2}\right)+
\hat{\Lambda}_5^2 \left(\frac{15 a^2}{16 \hat{c}_2^4}-\frac{27a \hat{c}_1^2}{8 \hat{c}_2^5}
+\frac{9 \hat{c}_1^4}{8 \hat{c}_2^6}\right)\nn\\
+\hat{\Lambda}_5^3 \left(\frac{45 a^2 \hat{c}_1}{8 \hat{c}_2^7}
-\frac{189a \hat{c}_1^3}{16 \hat{c}_2^8}+\frac{27 \hat{c}_1^5}{8 \hat{c}_2^9}\right)
+O(\hat{\Lambda}_5^4)\qquad
\eea
Plugging above expansions in (\ref{prep_exact})  we get the generalized 
prepotential as a series in $\hat{\Lambda}_5$. We have checked that the terms 
available from CFT calculation are in exact agreement with this series.

We end this section with one more remark. The theory considered 
in this paper actually is certain deformation of standard ${\cal H}_0$ Argyres-Douglass
theory discussed in literature (see e.g. the review \cite{Tachikawa:2013kta}). 
The proper ${\cal H}_0$ theory is obtained upon specialization
\bea
\label{spec_H0} 
\hat{c}_1=\frac{c_{AD}^{3/4}}{\sqrt{3}}\,;\qquad \hat{c}_2=\sqrt{\frac{3}{2}}\, c_{AD}^{1/4}\,;\qquad 
\hat{\Lambda}_5=\frac{\sqrt{2}}{4}\,;\qquad \hat{v}=u_{AD}
\eea 
where $c_{AD}$ and $u_{AD}$ are the standard conjugate to each other quantities of 
${\cal H}_0$ AD theory with scaling dimension $[{4\over 5}]$ and $[{6\over 5}]$ respectively.
It follows from (\ref{g2_g3_H0}) that in this special case the Weierstrass 
parameters simply coincide with $c_{AD}$ and $u_{AD}$: 
\be 
g_2=c_{AD}\,;\qquad g_3=u_{AD}
\ee
Specifying (\ref{prep_exact}) according to  (\ref{spec_H0}) and choosing 
$\epsilon_1=\epsilon_2$ (equivalently $s^2=4p$) one can easily check that 
our result reproduces formulae (4.3), (4.4) of \cite{Fucito:2023txg} derived 
just in this restricted setting.   
%%%%%%%%%%%%%%%%%%%%%%%%%%%%%%%%%%%%%%%%%%%%%%%%%%%%%%%%%%%%%%%%%%%%%%%%%%%%%%%%%
\section{NS limit and WKB analysis}
\label{NS_WKB}
%%%%%%%%%%%%%%%%%%%%%%%%%%%%%%%%%%%%%%%%%%%%%%%%%%%%%%%%%%%%%%%%%%%%%%%%%%%%%%%%%
In this section, using WKB method we investigate ${\cal H}_0$ AD theory in 
Nekrasov-Shatashvili (NS) limit $\epsilon_1\to 0$. Our approach is quite 
parallel to that of \cite{Fucito:2023plp} devoted to investigation of other AD theories. 
NS limit has attracted much attention due to its tight connection to 
quantum integrable systems \cite{Nekrasov:2009rc}. Direct application of localization 
technique in this limit leads to the concept of deformed (or quantum) SW curve  
\cite{Poghossian:2010pn,Fucito:2011pn} (see also \cite{Mironov:2009dv} for an earlier approach). 
Among other structures, Baxters's T-Q difference equation emerges quite naturally in 
this approach thus shedding new light on 2d/4d duality.  
By means of Fourier transform this T-Q equation immediately leads to a Schr\"{o}dinger-like 
equations with Plank's constant $\hbar=\epsilon_2$
\bea
\left(\hbar^2\frac{d^2}{dz^2}-\hat\phi_2(z)\right)\psi(z)=0
\label{schroedingereq}
\eea
The potential $\hat\phi_2(z)$ defines the SW-differential as in (\ref{SW_dif}). 
$\psi(z)$ can be interpreted as the partition function of certain quiver gauge theory, 
namely the AGT dual  \cite{Alday:2009aq} of 2d CFT conformal block with an extra 
degenerate field insertion.  

The results obtained in previous sections can be tested in NS limit $\epsilon_1 \to 0$ 
with small $\epsilon_2=\hbar$ using standard WKB ansatz:
\bea
\label{psi_ansatz}
\psi(z)=e^{\frac{1}{\hbar} {\cal F}(z)}
\eea
%%%%%%%%%%%%%%%%%%%%%%%%%%%%%%%%%%%%%%%%%%%%%%%%%%%%%%%%%%%%%%%%%%%%%%%%%%%%
\subsection{The quantum period}
Inserting (\ref{psi_ansatz}) into (\ref{schroedingereq}) for the "momentum"  
\bea
P(z)=\frac{{\cal F}'(z) }{\hbar}
\eea
we get the first order differential equation
\bea
P'(z)+P(z)^2-\frac{1}{\hbar^2}\hat\phi_2(z)=0 \label{difp}
\eea
Plugging semiclassical expansion
\bea
P(z)=\sum_{n=-1}^\infty\hbar^nP_n(z)
\eea
into (\ref{difp}) we get recursion relation
\bea
P_{n+1}(z)=\frac{1}{2\sqrt{\hat\phi_2(z)}}\left(
\frac{d}{dz}\,P_n(z)+\sum _{m=0}^n P_m(z) P_{n-m}(z)\right)
\eea
starting from
\bea
P_{-1}(z)=\sqrt{\hat\phi_2(z)}
\eea 
we can recursively  derive higher order terms $P_n(z)$, $n=0,1,2,\ldots $. 
One can show that all $P_n(z)$s with even $n$ are total derivatives, so 
that their integrals around closed cycles vanish. Thus only $P_n(z)$
with odd $n$ are relevant for  computation of the periods, and, eventually 
for the prepotential. Let us list $P_n(z)$ for $n=1,3,5$ explicitly
\bea
P_1(z)&=&-\frac{5 \hat\phi_2 '^2+4 \hat{\phi_2}  \hat\phi_2 ''}{32 \hat\phi_2 ^{5/2}} \\
P_3(z) &=&-\frac{221 \hat\phi_2 '^2 \hat\phi_2 ''}{256 \hat\phi_2 ^{9/2}}-
\frac{1105 \hat\phi_2 '^4}{2048 \hat\phi_2 ^{11/2}}-\frac{7 \hat\phi_2 ^{(3)} \hat\phi_2 '}{32 \hat\phi_2 ^{7/2}}
-\frac{19 \hat\phi_2 ''^2}{128 \hat\phi_2 ^{7/2}}+\frac{\hat\phi_2 ^{(4)}}{32 \hat\phi_2 ^{5/2}}\nonumber\\
P_5(z) &=&+\frac{248475 \hat\phi_2 '^4 \hat\phi_2 ''}{16384 \hat\phi_2 ^{15/2}}
-\frac{34503 \hat\phi_2 '^2 \hat\phi_2 ''^2}{4096 \hat\phi_2 ^{13/2}}
+\frac{1391 \hat\phi_2 ^{(3)} \hat\phi_2 ' \hat\phi_2 ''}{512 \hat\phi_2 ^{11/2}}-\frac{414125 \hat\phi_2 '^6}{65536 \hat\phi_2 ^{17/2}}
-\frac{1055 \hat\phi_2 ^{(3)} \hat\phi_2 '^3}{256 \hat\phi_2 ^{13/2}}
\nonumber\\
&+& \frac{815 \hat\phi_2 ^{(4)} \hat\phi_2 '^2}{1024 \hat\phi_2 ^{11/2}}-\frac{27 \hat\phi_2 ^{(5)} \hat\phi_2 '}{256 \hat\phi_2 ^{9/2}}
-\frac{55 \hat\phi_2 ^{(4)} \hat\phi_2 ''}{256 \hat\phi_2 ^{9/2}}
+\frac{631 \hat\phi_2 ''^3}{1024 \hat\phi_2 ^{11/2}}
-\frac{69 \left(\hat\phi_2^{(3)}\right)^2}{512 \hat\phi_2^{9/2}}+\frac{\hat\phi_2^{(6)}}{128 \hat\phi_2^{7/2}}\nonumber
\eea
These three expressions are sufficient
to calculate the $\epsilon$-corrections to the prepotential up to order $\epsilon_2^6$ included.
In case of our interest $\hat\phi_2$, (\ref{phi2h0}) is a function of a single quantum 
Coulomb branch parameter $\hat v$.  
The quantum $a$-period can be expanded as
\bea
\label{a_expansion_gen}
a(\hat{v})=a_0(\hat{v})+\epsilon_2^2a_2(\hat{v})+\epsilon_2^4a_4(\hat{v})+\epsilon_2^6a_6(\hat{v})+O(\epsilon_2^8)
\eea
with
\bea
\label{a_n_general}
a_n(\hat{v})=\oint_{\gamma_A} P_{n-1}(z)\frac{dz}{2\pi i}
\eea
Inverting above expansion one can represent $\hat{v}$ as a function of 
the flat coordinate $a$ as follows:
\bea
\hat{v}(a)=v_0(a)+\epsilon_2^2v_2(a)+\epsilon_2^4v_4(a)+\epsilon_2^6v_6(a)+\cdots
\label{uvsa}\eea
where the coefficient functions $v_n(a)$ can be uniquely determined by
inserting (\ref{uvsa}) in (\ref{a_expansion_gen}) and comparing two sides of the equality.
Here is what we get
\bea
\label{u_246_general}
&&v_2(a)= -\frac{a_2}{a_0'}\,\,; \quad
v_4(a)= -\frac{a_2{}^2 a_0''}{2 a_0'{}^3}
+\frac{a_2 a_2'}{a_0'{}^2}
-\frac{a_4}{a_0'}\\
&&v_6(a) = {-}\frac{a_2{}^3 a_0''{}^2}{2 a_0'{}^5}
{+}\frac{3 a_2{}^2 a_2' a_0''}{2 a_0'{}^4}-\frac{a_2{}^2 a_2''}{2 a_0'{}^3}{-}
\frac{a_4 a_2 a_0''}{a_0'{}^3}
{+}\frac{a_0{}^{(3)} a_2{}^3}{6 a_0'{}^4}
{+}\frac{a_2 a_4'}{a_0'{}^2}
{-}\frac{a_2 a_2'{}^2}{a_0'{}^3}
{+}\frac{a_4 a_2'}{a_0'{}^2}
{-}\frac{a_6}{a_0'}\nonumber
\eea
It is assumed that all the $a_n$'s and their derivatives on the r.h.s. 
are evaluated at the argument $v_0$ satisfying the equation $a_0(v_0)= a$.
%%%%%%%%%%%%%%%%%%%%%%%%%%%%%%%%%%%%%%%%%%%%%%%%%%%%%%%%%%%%%%%%%%%%%%%%%%%%%%%%%%%%%%%%%%%%%%%%%%%%%%%%%
\subsection{${\cal H}_0$ Argyres-Douglas theory in the NS limit}
It is straightforward to specialize above general scheme to the case 
of ${\cal H}_0$ theory which is characterized by $\hat{\phi}_2$ 
given in (\ref{phi2hat}).
As already mentioned for small $\hat{\Lambda}_5$ the $A$-cycle shrinks to 
a small contour around $z=0$ and the integrals (\ref{a_n_general}) can be 
computed by taking residues. We have expanded
the relevant quantities up to order $\hat{\Lambda}_5^6$ and computed 
$\epsilon_2$ corrections to $a$. Then using (\ref{u_246_general}) we have 
found $\hat{v}(a)$ up to order $\epsilon_2^6$. The results of
computations are presented in appendix \ref{WKB_results_h0}. It is also clarified 
there how to check these results against CFT. Explicit computations assure that the 
match is perfect.  NS limit of Argyres-Douglas theories has been addressed earlier in \cite{Ito:2018hwp}. 
	We have checked that the outcome of our elementary, perturbative in $\Lambda_5$, computations agree with the results of \cite{Ito:2018hwp}.
%%%%%%%%%%%%%%%%%%%%%%%%%%%%%%%%%%%%%%%%%%%%%%%%%%%%%%%%%%%%%%%%%%%%%%%%%%%%%%%%%%%%%%%%%%%%%%%%%%%%%%%%%
\section{The partition function of ${\cal H}_0$ Argyres-Douglas theory with 
$\epsilon_1=-\epsilon_2$ and Penlev\'{e} \RNum{1} $\tau$-function}
\label{painleve1}
%%%%%%%%%%%%%%%%%%%%%%%%%%%%%%%%%%%%%%%%%%%%%%%%%%%%%%%%%%%%%%%%%%%%%%%%%%%%%%%%%%%%%%%%%%%%%%%%%%%%%%%%%
The equation Penlev\'{e} \RNum{1} (shorthand notation P1)
\bea
\label{P1} 
q_{tt}=6q^2+t
\eea 
is the simplest among six second order ordinary differential equations in classification scheme
developed Painlev\'e and Gambier. 
The equation (\ref{P1}) can be represented in Hamiltonian form with time dependent Hamilton function 
\bea 
\sigma(t)=\frac{q_t^2}{2}-2q^3-qt
\eea
which due to (\ref{P1}) itself satisfies the equation
\bea 
\sigma_{tt}^2=2(\sigma-t\sigma_t)-4\sigma_t^3
\eea 
$\tau$-function of P1 is introduced through the relation
\bea 
\tau(t)=\frac{\sigma_t }{\sigma }
\eea 
According to the conjecture proposed in \cite{Bonelli:2016qwg,Lisovyy_2017} 
along the 5 rays in complex $t$-plane
$\arg t=\pi,\pm 3\pi/5, \pm \pi/5$ the function $\tau(t)$ admits the following series representation
\bea 
&&\tau(t)=x^{-\frac{1}{10}}\sum_{n\in \mathbb{Z}}e^{in\rho}{\cal G}(\nu+n,x )\,;\qquad 
24t^5+x^4=0\,,\,\,\,x\in \mathbb{R}_{\ge 0}\nn\\
&&{\cal G}(\nu ,x )=C(\nu ,x )\left[1+\sum_{k=1}^\infty \frac{D_k(\nu )}{x^k} \right]\nn\\
&&C(\nu ,x )=(2\pi)^{\frac{\nu}{2}}e^{\frac{x^2}{45}
+\frac{4}{5}i\nu x-\frac{i\pi \nu^2}{4}}x^{\frac{1}{12}
-\frac{\nu^2}{2}}48^{-\frac{\nu^2}{2}}G(1+\nu )
\eea
where $G(1+\nu )$ is Barnes $G$-function and the parameters $\nu$, $\rho$ are related 
to Stokes multipliers (see \cite{Lisovyy_2017}). The first three coefficients $D_k(\nu)$ 
explicitly read 
\bea 
&&D_1(\nu )=-\frac{i \nu (94\nu ^2+17)}{96}\nn\\
&&D_2(\nu )=-\frac{44180\nu ^6+170320\nu^ 4+74985\nu ^2+1344}{92160}\nn\\
&&D_3(\nu )=-\frac{i \nu(4152920 \nu ^8+45777060\nu ^6
+156847302\nu^4+124622833\nu ^2+13059000)}{26542080}\nn
\eea 
In analogy with previously known cases, it was anticipated that 
${\cal G}(\nu ,x )$ should be closely related to partition 
function of ${\cal H}_0$ theory in $\Omega$-background with $\epsilon_1=-\epsilon_2$.
Explicitly, under identification 
\bea 
x=\frac{2\left(c_2^4-6c_1c_2 \Lambda_5\right)^{5/4}}{3\Lambda_5^2}\,;\qquad \nu=-ia
\eea 
the quantity
\bea 
\log \left(1+\frac{D_1(\nu )}{x}+\frac{D_2(\nu )}{x^2}\right)+O(x^{-3})\nn
\eea
coincides with (\ref{logZcft}) incorporated  with terms coming from tree part (\ref{Zh0tree})   provided one sets $Q=0$.
Using holomorphic anomaly recursion we have computed prepotential  up to order 
$\Lambda_5^8$ which allowed not only to check the term $D_3$ but
also determines the next term 
\bea 
D_4(\nu)=\frac{4879681 \nu ^{12}}{127401984}+\frac{26452775 \nu ^{10}}{31850496}
+\frac{2887153423 \nu ^8}{424673280}+\frac{3126946955 \nu ^6}{127401984}\nn\\
+\frac{305174960717 \nu ^4}{10192158720}+
\frac{292259287 \nu ^2}{35389440}+\frac{49049}{460800}
\eea 
We have checked that the completely different computation based on conjecture by 
\cite{Bonelli:2016qwg,Lisovyy_2017}
gives exactly the same result.
%%%%%%%%%%%%%%%%%%%%%%%%%%%%%%%%%%%%%%%%%%%%%%%%%%%%%%%%%%%%%%%%%%%%%%%%%%%%%%%%%%%%%%%%%%%%%%%%%%%%%%%%%
\section{Summary}
\label{summary}
In this paper we have found a consistent way to define half integer rank irregular states. 
In particular the rank $5/2$ case, relevant for investigation of  ${\cal H}_0$  AD theory 
in $\Omega$-background is elaborated in full details (see (\ref{def_ir52}),  (\ref{ln_52})). 
We have conjectured that this state admits  expansion in terms of certain descendants 
of the rank $2$ irregular state (\ref{rank52_vs_rank2_expansion}).
Identifying the appropriate prefactor (\ref{factor_f}) we have computed the generalized 
descendants up to level 3 (see (\ref{desc_1}), (\ref{desc_2}) and (\ref{desc_3})).  
It is expected that also higher order terms can be fixed uniquely by imposing conditions (\ref{def_ir52}).
This result has been used to compute the  rank $5/2$ conformal block (\ref{logZcft}) which 
indeed, in case of vanishing $\Omega$-background, correctly reproduces SW curve result. 

We have exactly evaluated the periods of SW differential for (generalized) ${\cal H}_0$ theory 
(\ref{a}), (\ref{aD}). Then applying holomorphic anomaly recursion relation we find
$q$-exact prepotential up to $8$-th order in $\epsilon_{1,2}$. The result up 
to order $6$ are given in (\ref{prep_exact}). Order $\epsilon^8$ expressions are very large. 
They are available upon request.
  
Results for Nekrasov-Shatashvili limit obtained from WKB computations are presented 
in appendix \ref{WKB_results_h0}.

In section \ref{painleve1} we have shown that in restricted $\Omega$-background 
with $\epsilon_1+\epsilon_2=0$ both irregular state and holomorphic anomaly results 
fully agree with large time Painlev\'{e} \RNum{1} $\tau$-function expansion.

%%%%%%%%%%%%%%%%%%%%%%%%%%%%%%%%%%%%%%%%%%%%%%%%%%%%%%%%%%%%%%%%%%%%%%%%%%%%%%%%%%%
\acknowledgments
R.P. thanks F.Fucito and J.F.Morales for many useful discussions. 
His work has been supported by the Armenian SCS
grant 21AG-1C062. H.P. acknowledges support in the framework of 
Armenian SCS grants 21AG-1C060 and 20TTWS1C035.

%%%%%%%%%%%%%%%%%%%%%%%%%%%%%%%%%%%%%%%%%%%%%%%%%%%%%%%%%%%%%%%%%%%%%%%c
%\end{appendix}

  \begin{appendix}
%%%%%%%%%%%%%%%%%%%%%%%%%%%%%%%%%%%%%%%%%%%%%%%%%%%%%%%%%%%%%%%%%%%%%%%%%%
\section{Prepotential from irregular states}
%%%%%%%%%%%%%%%%%%%%%%%%%%%%%%%%%%%%%%%%%%%%%%%%%%%%%%%%%%%%%%%%%%%%%%%%%%%%%%%%%%%%%%%
\subsection{Expansion of the rank $5\over 2$ irregular state in terms of rank $2$ 
	states}
\label{rank2_desc}
It is useful to modify slightly the Virasoro generator $L_{-1}$ denoting 
\bea 
\mathbb{L}_{-1}=L_{-1}-5c_{1}\partial_{c_2}\,; \qquad \mathbb{L}_n=L_{n}\quad \text{if} \,\,\,
n\neq -1 \nn
\eea
Then the level-$k$ descendant can be represented as 
\footnote{Here for a partition $Y=\{Y_1,Y_2,\cdots ,Y_r\}$, $Y_1\ge Y_2\ge \cdots \ge Y_r$, 
		we denote $\mathbb{L}_{-Y}\equiv \mathbb{L}_{-Y_1}\cdots \mathbb{L}_{-Y_r} $.}
\bea
|I^{(2)}_k(c_0,c_1,c_2;\beta_0,\beta_1 ) \rangle=\sum_{Y,n,m}d_{Y,n,m}\mathbb{L}_{-Y}\partial_{c_1}^{n}
\partial_{c_2}^{m}|I^{(2)}(c_0,c_1,c_2;\beta_0,\beta_1 ) \rangle
\eea 
with some coefficients $d_{Y,n,m}$, which can be recursively determined from irregular state conditions 
(\ref{def_ir52}), (\ref{ln_52}).

Explicitly for level-$2$ case we get (naturally only non-zero coefficients are displayed)
\footnote{Below we have shifted $c_0\to c_0+\frac{3Q}{2}$ to make 
formulae a bit more compact.}:
\begin{tiny} 
	\bea
	\label{desc_2} 
	&&d_{\{\},00}=\frac{121 c_1^6 \left(2 c_0+Q\right){}^2}{288 c_2^8}+
	\frac{c_1^4 \left(2 c_0+Q\right) \left(11 \left(-12 c_0 Q+16 c_0^2+Q^2\right)-266\right)}{96 c_2^7}\nn\\
	&&+\frac{c_1^2 \left(8 c_0 \left(74-3 Q^2\right) Q+16 c_0^2 \left(11 Q^2-42\right)-384 c_0^3 Q+256 c_0^4+Q^4-86 Q^2+49\right)}{128 c_2^6}+\frac{8 c_0 \left(-45 c_0 Q+32 c_0^2+16 Q^2-8\right)+19 Q}{384 c_2^5}\nn\\
	&&d_{\{\},01}=\frac{c_1^2}{96 c_2^5}+\frac{103 c_0}{144 c_2^4}\nn\\
	&&d_{\{\},10}=-\frac{11 c_1^5 \left(2 c_0+Q\right)}{24 c_2^7}
	+\frac{c_1^3 \left(196 c_0 Q+32 c_0^2-9 Q^2+153\right)}{144 c_2^6}+\frac{c_1 \left(48 c_0 \left(-12 c_0 Q+16 c_0^2+Q^2\right)-802 c_0+57 Q\right)}{576 c_2^5}\nn\\
	&&d_{\{\},20}=\frac{c_1^4}{8 c_2^6}-\frac{c_0 c_1^2}{3 c_2^5}+\frac{256 c_0^2-57}{1152 c_2^4}\,;\qquad
	d_{\{1\},00}=\frac{c_1 \left(72 c_0 Q-96 c_0^2-6 Q^2+95\right)}{288 c_2^5}-\frac{11 c_1^3 \left(2 c_0+Q\right)}{72 c_2^6}\nn\\
	&&d_{\{1\},10}=\frac{c_1^2}{12 c_2^5}-\frac{c_0}{9 c_2^4}\,;\qquad
	d_{\{1,1\},00}=\frac{1}{72 c_2^4}\,;\qquad d_{\{2\},00}=-\frac{7}{96 c_2^4}
	\eea
\end{tiny}
For level-$3$ coefficients we get:
\begin{tiny}
	\bea
	\label{desc_3} 
	&&d_{\{\},00}=-\frac{1331 c_1^9 \left(2 c_0+Q\right){}^3}{10368 c_2^{12}}
	-\frac{11 c_1^7 \left(2 c_0+Q\right){}^2 \left(11 \left(-12 c_0 Q+16 c_0^2+Q^2\right)
		-521\right)}{2304 c_2^{11}}\nn\\
	&&-\frac{c_1^5 \left(2 c_0+Q\right)}{7680 c_2^{10}}
	\left(5 \left(-264 c_0 Q^3+16 \left(121 c_0^2-91\right) Q^2 
	+8 c_0 \left(1579-528 c_0^2\right) Q+64 c_0^2 \left(44 c_0^2-243\right)
	+11 Q^4\right)+76981\right)\nn\\
	&&+\frac{c_1^3}{9216 c_2^9}\left(160 c_0^3 \left(54 Q^2-901\right) Q
	+4 c_0 \left(27 Q^4-4304 Q^2+43064\right) Q+1280 c_0^4 \left(65-18 Q^2\right)\right.\nn\\
	&&\left.-16 c_0^2 \left(90 Q^4-5192 Q^2+10801\right)
	+27648 c_0^5 Q-12288 c_0^6-3 Q^6+765 Q^4-32335 Q^2+14259
	\right)\nn\\
	&&+\frac{c_1}{3072 c_2^8}
	\left(32 c_0^3 \left(424-207 Q^2\right)+8 c_0^2 Q \left(237 Q^2-2591\right)
	-4 c_0 \left(32 Q^4-2217 Q^2+880\right)+8832 c_0^4 Q-4096 c_0^5
	+7 Q \left(193-85 Q^2\right)\right)\nn
	\\
	&&d_{\{\},01}=-\frac{c_1^3 \left(22120 c_0 Q+46040 c_0^2+45 Q^2-14901\right)}{34560 c_2^8}-\frac{11 c_1^5 \left(2 c_0+Q\right)}{1152 c_2^9}\nn\\
	&&+\frac{c_1 \left(-103 c_0 \left(15 \left(-12 c_0 Q+16 c_0^2+Q^2\right)-343\right)-2336 Q\right)}{17280 c_2^7}\nn\\
	&&d_{\{\},10}=\frac{c_1^6 \left(2 c_0+Q\right) \left(-1672 c_0 Q+616 c_0^2+99 Q^2-3978\right)}{1728 c_2^{10}}+\frac{121 c_1^8 \left(2 c_0+Q\right){}^2}{576 c_2^{11}}\nn\\
	&&+\frac{c_1^4 \left(4 c_0^2 \left(2508 Q^2+2539\right)+8 c_0 Q \left(5770-147 Q^2\right)-6144 c_0^3 Q-9984 c_0^4+3 \left(9 Q^4-1271 Q^2+7353\right)\right)}{6912 c_2^9}\nn\\
	&&+\frac{c_1^2 \left(480 c_0^3 \left(929-132 Q^2\right)+360 c_0^2 Q \left(24 Q^2-1097\right)+c_0 \left(-360 Q^4+64050 Q^2-347414\right)+138240 c_0^4 Q-92160 c_0^5-855 Q^3+43269 Q\right)}{69120 c_2^8}\nn\\
	&&+\frac{4 c_0^2 \left(5791-1440 Q^2\right)+16200 c_0^3 Q-6069 c_0 Q-11520 c_0^4+1620 \left(3 Q^2-1\right)}{25920 c_2^7}\nn\\
	&&d_{\{\},11}=\frac{c_1^4}{192 c_2^8}+\frac{101 c_0 c_1^2}{288 c_2^7}+\frac{146-515 c_0^2}{1080 c_2^6}\nn\\
	&&d_{\{\},20}=\frac{c_1^5 \left(284 c_0 Q+208 c_0^2-9 Q^2+297\right)}{576 c_2^9}+\frac{c_1^3 \left(2 c_0 \left(32 \left(-152 c_0 Q+56 c_0^2+9 Q^2\right)-8793\right)+1311 Q\right)}{13824 c_2^8}-\frac{11 c_1^7 \left(2 c_0+Q\right)}{96 c_2^{10}}\nn\\
	&&+\frac{c_1 \left(15 \left(57-256 c_0^2\right) Q^2+60 c_0 \left(768 c_0^2-323\right) Q+80 c_0^2 \left(1727-768 c_0^2\right)-27767\right)}{138240 c_2^7}\nn
	\eea
	\bea
	&&d_{\{\},30}=\frac{c_1^6}{48 c_2^9}-\frac{c_0 c_1^4}{12 c_2^8}
	+\frac{\left(256 c_0^2-57\right) c_1^2}{2304 c_2^7}+\frac{c_0 \left(171-256 c_0^2\right)}{5184 c_2^6}\nn\\
	&&d_{\{1\},00}=\frac{121 c_1^6 \left(2 c_0+Q\right){}^2}{1728 c_2^{10}}+\frac{c_1^4 \left(2 c_0+Q\right) \left(66 \left(-12 c_0 Q+16 c_0^2+Q^2\right)-2575\right)}{3456 c_2^9}\nn\\
	&&+\frac{c_1^2 \left(15 \left(-72 c_0 Q^3+\left(528 c_0^2-347\right) Q^2+36 c_0 \left(79-32 c_0^2\right) Q+16 c_0^2 \left(48 c_0^2-215\right)+3 Q^4\right)+38932\right)}{34560 c_2^8}\nn\\
	&&+\frac{8 c_0 \left(45 \left(-45 c_0 Q+32 c_0^2+16 Q^2\right)-4022\right)+855 Q}{103680 c_2^7}\nn\\
	&&d_{\{1,1\},00}=-\frac{11 c_1^3 \left(2 c_0+Q\right)}{864 c_2^8}
	+\frac{c_1 \left(36 c_0 Q-48 c_0^2-3 Q^2+92\right)}{1728 c_2^7}\nn\\ 
	&&d_{\{1\},10}=-\frac{11 c_1^5 \left(2 c_0+Q\right)}{144 c_2^9}+
	\frac{c_1^3 \left(392 c_0 Q+64 c_0^2-18 Q^2+573\right)}{1728 c_2^8}+
	\frac{c_1 \left(2 c_0 \left(24 \left(-12 c_0 Q+16 c_0^2+Q^2\right)-757\right)+57 Q\right)}{3456 c_2^7}\nn\\ 
	&&d_{\{2\},00}=\frac{77 c_1^3 \left(2 c_0+Q\right)}{1152 c_2^8}+
	\frac{7 c_1 \left(45 \left(-12 c_0 Q+16 c_0^2+Q^2\right)-1421\right)}{34560 c_2^7};\qquad 
	d_{\{1,1\},10}=\frac{c_1^2}{144 c_2^7}-\frac{c_0}{108 c_2^6}\nn\\ 
	&&d_{\{2\},10}=\frac{7 c_0}{144 c_2^6}-\frac{7 c_1^2}{192 c_2^7}\,;\qquad
	d_{\{1\},01}=\frac{c_1^2}{576 c_2^7}+\frac{103 c_0}{864 c_2^6}\nn\\
	&&d_{\{1\},20}=\frac{c_1^4}{48 c_2^8}-\frac{c_0 c_1^2}{18 c_2^7}+\frac{256 c_0^2-57}{6912 c_2^6}\,;\qquad
	d_{\{3\},00}=\frac{343}{8640 c_2^6}\,;\qquad
	d_{\{2,1\},00}=-\frac{7}{576 c_2^6}\,;\qquad
	d_{\{1,1,1\},00}=\frac{1}{1296 c_2^6}
	\eea 
\end{tiny}
We have calculated also the level $4$ term, but it is too lengthy to be 
displayed here. The authors will be glad to make this expression available upon request.
%%%%%%%%%%%%%%%%%%%%%%%%%%%%%%%%%%%%%%%%%%%%%%%%%%%%%%%%%%%%%%%%%%%%%%%%%%%%%%%%%%%%%%%%%%%%%%%
\subsection{The irregular conformal block}
\label{irr_block_4}
Now it is straightforward to calculate the matrix element (\ref{ZH0_def})
up to order $O(\Lambda_5^4)$. After factoring out the tree part (\ref{Zh0tree}) we get
\begin{small}
\bea
\label{logZcft}
\log Z_{ {\cal H}_0 \rm inst}&=&
-\frac{c_1 \left(7 Q^2+60 a^2-2\right)}{16 c_2^3}\, \Lambda _5+
\left(\frac{a \left(77 Q^2+188 a^2-34\right)}{128 c_2^5}
-\frac{3 c_1^2 \left(7 Q^2+60 a^2-2\right)}{16 c_2^6}\right)\Lambda _5^2 \nn\\
&+&\left(\frac{15 c_1 a \left(77 Q^2+188 a^2-34\right)}{256 c_2^8}
-\frac{3 c_1^3 \left(7 Q^2+60 a^2-2\right)}{4 c_2^9}\right)\Lambda _5^3\nn\\
&+&\left(\frac{1}{c_2^{10}}\left(-\frac{101479 Q^4}{491520}+\frac{32179 Q^2}{122880}
-\frac{21}{640}
+\left(\frac{3677}{2048}-\frac{13937 Q^2}{4096}\right) a^2-\frac{7717 a^4}{2048}\right)\right. \nn\\
&&+\left.\frac{c_1^2}{c_2^{11}}\left(\frac{405 \left(77 Q^2-34\right) a}{1024}+\frac{19035 a^3}{256}\right)
+\frac{c_1^4}{c_2^{12}}\left(-\frac{189 Q^2}{8}-\frac{405 a^2}{2}+\frac{27}{4}\right)\right)\Lambda _5^4
+\cdots\nn\\
\eea  
where
\be
\label{t_c0} 
a=c_0-\frac{3Q}{2}
\ee
Using (\ref{G_Matone_CFT}) and (\ref{Zh0tree}) for the Coulomb branch modulus we obtain
\bea
\label{vcft}
v=&-&\frac{c_1\left(7 Q^2+60 a^2-2\right) \Lambda _5 }{16 c_2^3}+
\left(\frac{a \left(77 Q^2+188 a^2-34\right)}{128 c_2^5}
-\frac{3 c_1^2 \left(7 Q^2+60 a^2-2\right)}{16 c_2^6}\right)\Lambda _5^2\nn\\
&+&\left(\frac{15 c_1 a \left(77 Q^2+188 a^2-34\right)}{256 c_2^8}
-\frac{3 c_1^3 \left(7 Q^2+60 a^2-2\right)}{4 c_2^9}\right)\Lambda _5^3 +O(\Lambda_5^4)
\eea 
\end{small}
%%%%%%%%%%%%%%%%%%%%%%%%%%%%%%%%%%%%%%%%%%%%%%%%%%%%%%%%%%%%%%%%%%%%%%%%%%%%%%%%%%%%%%%%%%%%%%%%%%%%%%%%%%%%%
%%%%%%%%%%%%%%%%%%%%%%%%%%%%%%%%%%%%%%%%%%%%%%%%%%%%%%%%%%%%%%%%%%%%%%%%%%%%%%%%%%%%%%%%%%%%%%%%%%%%%%%%%%%%%
\section{NS limit}
\label{WKB_results_h0}
The quadratic differential $\hat{\phi}_2dz^2$ in this case is given by (\ref{phi2hat}).
The integrals (\ref{a_n_general}) in the small $\hat{\Lambda}_5$ 
limit can be computed by taking residues at $z=0$. For un-deformed $a_0(\hat{v})$ we get
\begin{small}
\bea
\label{a0}
a_0(\hat v)&=&\frac{2 \hat{v}-\hat{c}_1^2}{2 \hat{c}_2}+\frac{\hat{c}_1 \hat{\Lambda} _5 \left(6 \hat{v}-5 \hat{c}_1^2\right)}{4 \hat{c}_2^4}
-\frac{15 \hat{\Lambda} _5^2 \left(-28 \hat{c}_1^2 \hat{v}+21 \hat{c}_1^4+4 \hat{v}^2\right)}{64 \hat{c}_2^7}\nn\\
&-&\frac{21 \hat{c}_1 \hat{\Lambda} _5^3 \left(-220 \hat{c}_1^2 \hat{v}+143 \hat{c}_1^4+60 \hat{v}^2\right)}{128 \hat{c}_2^{10}}
+\frac{1155 \hat{\Lambda} _5^4 \left(-156 \hat{c}_1^2 \hat{v}^2+390 \hat{c}_1^4 \hat{v}-221 \hat{c}_1^6+8 \hat{v}^3\right)}{2048 \hat{c}_2^{13}}\nn\\
&+&\frac{9009 \hat{c}_1 \hat{\Lambda} _5^5 \left(-340 \hat{c}_1^2 \hat{v}^2+646 \hat{c}_1^4 \hat{v}-323 \hat{c}_1^6+40 \hat{v}^3\right)}{4096 \hat{c}_2^{16}}\nn\\
&-&\frac{51051 \hat{\Lambda} _5^6 \left(15960 \hat{c}_1^4 \hat{v}^2-3040 \hat{c}_1^2 \hat{v}^3-24472 \hat{c}_1^6 \hat{v}
+10925 \hat{c}_1^8+80 \hat{v}^4\right)}{131072 \hat{c}_2^{19}}+O(\hat{\Lambda}_5^7)
\eea
\end{small}
Similarly for $A$-cycle corrections $a_{2,4,6}(\hat{v})$ we have obtained
\begin{small}
\bea
a_2(\hat{v})&=&-\frac{7 \hat{\Lambda} _5^2}{64 \hat{c}_2^5}-\frac{105 \hat{c}_1 \hat{\Lambda} _5^3}{128 \hat{c}_2^8}
+\frac{105 \hat{\Lambda} _5^4 \left(26 \hat{v}-121 \hat{c}_1^2\right)}{2048 \hat{c}_2^{11}}
-\frac{15015 \hat{c}_1 \hat{\Lambda} _5^5 \left(13 \hat{c}_1^2-6 \hat{v}\right)}{4096 \hat{c}_2^{14}}\nn\\
&-&\frac{15015 \hat{\Lambda} _5^6 \left(-1156 \hat{c}_1^2 \hat{v}+1615 \hat{c}_1^4+76 \hat{v}^2\right)}{65536 \hat{c}_2^{17}}
+O(\hat{\Lambda}_5^7)\nn\\
a_4(\hat{v})&=&-\frac{119119 \hat{\Lambda} _5^6}{131072 \hat{c}_2^{15}}+O(\hat{\Lambda}_5^7)\,;\qquad
a_6(\hat{v})=O(\hat{\Lambda}_5^7)
\eea
\end{small}
Inverting series $a_0(\hat{v})$ (\ref{a0}) for un-deformed modulus $\hat{v}_0(a)$ we get
\begin{small}
\bea
\hat{v}_0(a)&=&\frac{\hat{c}_1^2}{2}+\hat{c}_2 a+\left(\frac{\hat{c}_1^3}{2 \hat{c}_2^3}-\frac{3 \hat{c}_1 a}{2 \hat{c}_2^2}\right)\hat{\Lambda} _5 
+\left(\frac{15 a^2}{16 \hat{c}_2^4}-\frac{27 \hat{c}_1^2 a}{8 \hat{c}_2^5}+\frac{9 \hat{c}_1^4}{8 \hat{c}_2^6}\right)\hat{\Lambda} _5^2 \nn\\
&+&\left(\frac{45 \hat{c}_1 a^2}{8 \hat{c}_2^7}-\frac{189 \hat{c}_1^3 a}{16 \hat{c}_2^8}+\frac{27 \hat{c}_1^5}{8 \hat{c}_2^9}\right)\hat{\Lambda} _5^3 
+\left(\frac{135 \hat{c}_1^2 a^2}{4 \hat{c}_2^{10}}-\frac{705 a^3}{256 \hat{c}_2^9}
-\frac{6237 \hat{c}_1^4 a}{128 \hat{c}_2^{11}}+\frac{189 \hat{c}_1^6}{16 \hat{c}_2^{12}}\right)\hat{\Lambda} _5^4\nn\\
&+&\left(\frac{405 \hat{c}_1^3 a^2}{2 \hat{c}_2^{13}}-\frac{19035 \hat{c}_1 a^3}{512 \hat{c}_2^{12}}
-\frac{56133 \hat{c}_1^5 a}{256 \hat{c}_2^{14}}+\frac{729 \hat{c}_1^7}{16 \hat{c}_2^{15}}\right)\hat{\Lambda} _5^5\nn\\
&+&\left(\frac{1215 \hat{c}_1^4 a^2}{\hat{c}_2^{16}}
-\frac{742365 \hat{c}_1^2 a^3}{2048 \hat{c}_2^{15}}+\frac{115755 a^4}{8192 \hat{c}_2^{14}}
-\frac{1066527 \hat{c}_1^6 a}{1024 \hat{c}_2^{17}}+\frac{24057 \hat{c}_1^8}{128 \hat{c}_2^{18}}\right)\hat{\Lambda} _5^6 
+O(\hat{\Lambda}_5^7)
\eea
\end{small}
Using formulae  (\ref{u_246_general}), for the $\epsilon_2$ corrections $ \hat{v}_{2,4,6}$ we obtain
\begin{small}
\bea
\hat{v}_2(a)&=&\frac{7 \hat{\Lambda} _5^2}{64 \hat{c}_2^4}+\frac{21 \hat{c}_1 \hat{\Lambda} _5^3}{32 \hat{c}_2^7}+
\left(\frac{63 \hat{c}_1^2}{16 \hat{c}_2^{10}}-\frac{1155 a}{1024 \hat{c}_2^9}\right)\hat{\Lambda} _5^4\nn\\
&+&\left(\frac{189 \hat{c}_1^3}{8 \hat{c}_2^{13}}-\frac{31185 \hat{c}_1 a}{2048 \hat{c}_2^{12}}\right)\hat{\Lambda} _5^5
+\hat{\Lambda} _5^6 \left(\frac{209055 a^2}{16384 \hat{c}_2^{14}}
-\frac{1216215 \hat{c}_1^2 a}{8192 \hat{c}_2^{15}}+\frac{567 \hat{c}_1^4}{4 \hat{c}_2^{16}}\right)+O(\hat{\Lambda}_5^7)\nn\\
\hat{v}_4(a)&=&\frac{101479 \hat{\Lambda} _5^6}{131072 \hat{c}_2^{14}}+O(\hat{\Lambda}_5^7)\,;\qquad 
\hat{v}_6(a)=O(\hat{\Lambda}_5^7)
\eea
Now one can easily check that 
\[
\hat{v}_0(a)+\epsilon_2^2\hat{v}_2(a)+\cdots 
\]
is in complete agreement with the result obtained by applying 2d CFT/AGT 
map (\ref{map_CFT_H0}) to (\ref{vcft}) and setting $\epsilon_1=0$.  
\end{small}
%%%%%%%%%%%%%%%%%%%%%%%%%%%%%%%%%%%%%%%%%%%%%%%%%%%%%%%%%%%%%%%%%%%%%%%%%%%%%%%%%%%%%%%%%%%%%%%%%%%%%%%%%%%%%
\end{appendix}

\bibliographystyle{JHEP}
%\bibliography{references}
\providecommand{\href}[2]{#2}\begingroup\raggedright\endgroup
\end{document}